\documentclass[12pt,a4paper]{article}
\usepackage[utf8]{inputenc}
\usepackage[portuguese]{babel}
\usepackage[T1]{fontenc}
\usepackage{amsmath}
\usepackage{amsfonts}
\usepackage{amssymb}
\usepackage{graphicx}
\usepackage[left=2cm,right=2cm,top=2cm,bottom=2cm]{geometry}
\author{L. B. A. Mélo, Daniel Felinto, and Marcio H. G. de Miranda\thanks{marcio.miranda@ufpe.br} \and \\ \textit{Departamento de Física, Universidade Federal de Pernambuco,} \\ \textit{Recife, Pernambuco, Brazil} \and \\ \small{\textbf{This work was previously published on JOSA B \cite{Melo:23}}}}
\title{Optimized Phase Masks for Absorption of Ultra-Broadband Pulses by Narrowband Atomic Ensembles}
\date{}
\begin{document}

\maketitle

\newpage

\section*{Abstract}

By combining genetic algorithm and a spatial light modulator we theoretically analyse how to improve a two-photon cascade absorption in atomic ensembles, inspecting the impact of various configurations and parameters in the optimized phase mask. At low atomic densities, we compare the cases of sequential transitions with the two photons coming from the same pulse or from two different pulses. For the former, we predict an enhancement by a factor of $9.5$, similar to what was previously reported in the literature [Phys. Rev. Lett. {\bf 86}, 47 (2002)]. For the later, on the other hand, we obtain an enhancement factor of $26$ times. This absorption of two photons by different pulses is of particular interest for the storage of ultra-broadband single photons by atomic ensembles, in which case the second photon would come from a control pulse. We investigate this process as a function of the atomic density, demonstrating enhancements by factors up to 3 for the two-photon absorption after propagating through large optical depths. However, for the experimental conditions considered in the previous work by Carvalho {et al.} [Phys. Rev. A {\bf 101}, 053426 (2020)], in terms of control power and optical depths, we show that this enhancement in two-photon absorption would still result in just a modest increase of the absorption of a weak probe pulse.

\section{Introduction}

The problem of storing information from ultra-broadband light pulses in atomic memories with narrow bandwidth has witnessed significant progress in the last decade \cite{Bustard2013,Bustard2015,England2015,SHINBROUGH23}. This approach holds the potential to solve the problem of scalability in applications of photons resulting from Spontaneous Parametric Down Conversion (SPDC) \cite{carvalho2020enhanced}, being also a possible way to connect SPDC photons from satellites~\cite{yin2017satellite} to local quantum networks in ground stations. Bearing in mind these applications for the storage of SPDC photons, Carvalho {\it et al.} proposed and demonstrated how a two-photon cascade transition could be employed as an alternative to a more common $\Lambda$ transition for the absorption of a weak ultra-short pulse by a narrow-band dense atomic ensemble \cite{carvalho2020enhanced}. In the cascade transition, the weak field would act on the lower transition and the two-photon condition would be fulfilled by a stronger control field in the upper transition. However, the absorption enhancement reported in Ref.~\cite{carvalho2020enhanced}, when compared to simple linear absorption, was quite small, on the order of 0.3\%. A pathway for improvement was discussed in Ref.~\cite{carvalho2020enhanced}, including optimizations of optical depth and control power. Here we explore another optimization route to enhance absorption: the use of phase masks to shape the optical pulses. In this process, we revisit the old problem of coherent control of two-photon transitions by phase masks~\cite{Meshulach1999,dudovich2001transform}. We show that the sequential transition proposed for photon storage in Ref.~\cite{carvalho2020enhanced} is better suited for optimization, leading to higher enhancement factors, than the most well known results for sequential transitions by single pulses~\cite{dudovich2001transform}.

The possibility of increasing a resonant two-photon cascade absorption (TPCA) through spectral phase adjustments was demonstrated back in 2001 by Dudovich et al. \cite{dudovich2001transform}. The idea came from analyzing a basic TPCA modeling equation, where a destructive interference was identified when using a Fourier Transform Limited (FTL) pulse in the transition. With a clever spectral phase modification, Dudovich et al. achieved an improvement by a factor of 7 in the fluorescence from the most excited state. This result found many applications~\cite{silberberg2009quantum} and was developed in various directions along the years with the increasing use of adaptive algorithms~\cite{lozovoy2003multiphoton,ballard2002simultaneous,comstock2004multiphoton,roslund2009gradient, ballard2002optimization, ando2002optimization, deng2019tuning, huang2019optimizing}. In our work, we will show how one can improve a TPCA by using two pulsed lasers, each exciting a different transition. When it comes to adaptive coherent control, there is an advantage of investigating the optimal solution of a complex multi-parameter problem without the need of testing all the topological possibilities. Among the adaptive techniques, the Genetic Algorithm (GA) has the characteristic of finding the optimum global solution with high probability and it is no strange to spectral phase optimization \cite{deng2019tuning, huang2019optimizing}. By combining these two ideas, we predict enhancements by factors around 20 in the TPCA for a single atom, when compared to excitation by two different FTL pulses.

In this work, our main objective is to bring these developments for coherent control of TPCA to the problem of ultra-broadband photon storage by dense atomic ensembles. The investigation in Ref.~\cite{carvalho2020enhanced} focused on the perturbative regime with respect to control power, since this corresponded to their experimental conditions. In these circumstances, the maximum enhancement of absorption due to the control pulse was quite small, but measurable. The above discussion about optimization of TPCA by phase masks would already apply to this problem at lower densities and higher control power. However, for any upper limit of control power, it is possible to enhance absorption by increasing the density of atoms in the sample. This pathway has to take into account the back action of the medium distorting the weak pulse in a way that decreases considerably its single-photon absorption, forming a so-called zero-area pulse~\cite{crisp1970propagation,Rothenberg1984,costanzo2016zero}. The distortion of the weak pulse decreases the enhancement factor obtained from the optimized phase masks. Even though, we were able to double the TPCA for optical depths (OD) as high as 720. This indicates that the pathway involving simultaneously enhancement of control power and number of atoms may still be optimized by phase masks. However, as we optimize the phase masks directly into the theoretical analysis of Ref.~\cite{carvalho2020enhanced} considering the reported experimental conditions, we obtain only a much more modest increase of around 50\%.

In the following, Sec.~\ref{SECSilb} demonstrates how the GA optimizes a TPCA from a single pulse. In Sec.~\ref{SEC ThinSample}, the same approach is applied to a system with two different laser pulses acting upon a single atom, bringing the simulation closer to the goal of this work. Also in this section, we introduce the effect over TPCA of a weak pulse with zero-area envelope, keeping the control as a FTL pulse. In Sec.~\ref{SECEnhanced}, we use the previous results to understand the phase-mask optimization for TPCA in a dense atomic medium, for different optical depths, under the conditions of Ref.~\cite{carvalho2020enhanced}. Finally, in Sec.~\ref{conclusion} we draw our conclusions.

\section{Phase masks for two-photon cascade absorption}
\label{SECSilb}

Following Ref.~\cite{dudovich2001transform}, we begin by employing a resonant TPCA model in Secs.~\ref{SECSilb} and~\ref{SEC ThinSample}. Consider a three-level system, with $|1\rangle$, $|2\rangle$, and $|3\rangle$, respectively, as the lowest, intermediate, and most excited level. The interaction potential under the electric dipole approximation and in the interaction representation is given by

\begin{eqnarray}
    \hat{V}_I(t) = - \mu_{12} e^{- i \omega_{12} t } E_l(t) | 1 \rangle \langle 2 | - \mu_{23} e^{- i \omega_{23} t } E_u(t - \tau) | 2 \rangle \langle 3 | + h.c. .
\end{eqnarray}

\noindent where $\mu_{ij}$ is the electric dipole for the $i \longrightarrow j$ transition, $E_l$ is the field acting upon the lower transition, $E_u$ the field acting on the upper transition and $\tau$ the delay between the fields. $\omega_{ij}$ are the frequencies for the transitions $i\rightarrow j$.

The probability amplitude for level $|i\rangle$ is given by $c_i$. In first order of approximation, $c_3(t)$ is then given by

\begin{eqnarray}
	c_3(t) \approx - \frac{\mu_{12} \mu_{23}}{\hbar^2} \int_{-\infty}^t \int_{- \infty}^{t'} dt' dt" e^{ i \omega_{12} t' } e^{ i \omega_{23} t'' } E_u^*(t' - \tau) E_l^*(t''). 
\end{eqnarray}

\noindent Considering $E_k(t) = \mathcal{E}_k(t) e^{i \omega_k t} + c.c.$, where $\mathcal{E}_k(t)$ and $\omega_k$ are the slow envelope and center frequency of field $k$, with $k=u$ or $l$, the rotating wave approximation leads to

\begin{eqnarray}
    c_3(t) \approx - \frac{\mu_{12} \mu_{23}}{\hbar^2} e^{i \omega_{23} \tau} \int_{-\infty}^{t - \tau} \int_{- \infty}^{t' + \tau} e^{ i (\omega_{23} - \omega_u) t' }  e^{ i (\omega_{12} - \omega_l) t'' } \mathcal{E}_u^*(t') \mathcal{E}_l^*(t'') dt' dt'', \label{EQU Base}
\end{eqnarray}

\noindent which will be the basic modelling equation for this section and the next.

\subsection{Phase-Matching Problem}

\begin{figure}[htb]
    \center
    \includegraphics[width = 0.7\linewidth]{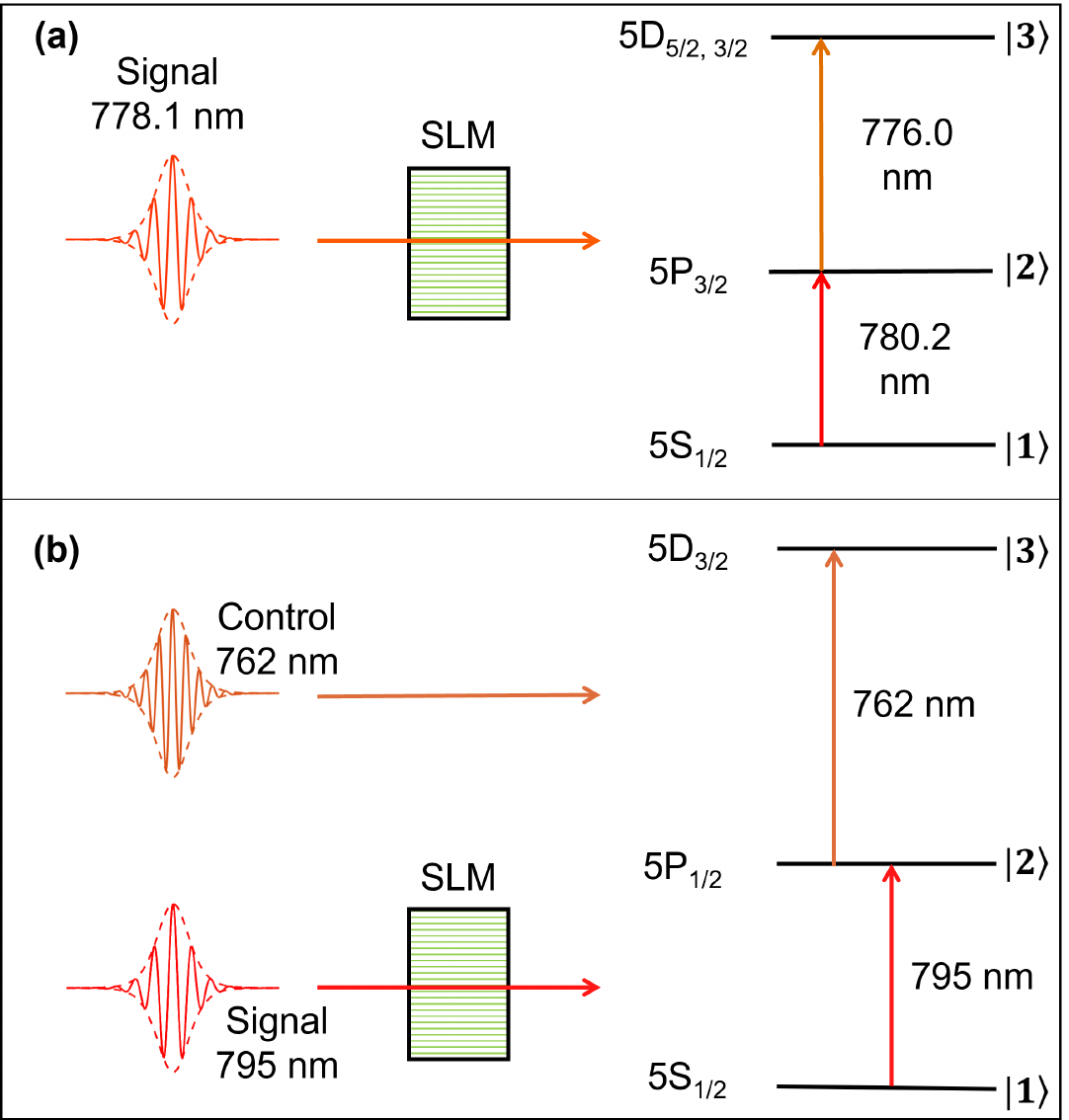}
    \caption{(a) Two-photon cascade absorption by a single laser in rubidium atoms. The laser, centered at 778.1 nm with 18 nm bandwidth, excite both 5S-5P (780.2 nm) and 5P-5D (776.0 nm) transitions. (b) Two-photon cascade absorption by two laser pulses. Each laser is resonant with only one transition and the signal pulse alone goes through the SLM to be optimized. The control pulse is centered at 762 nm with 10.4 nm bandwidth, and the signal pulse is centered at 795 nm with 7.5 nm bandwidth.} 
    \label{f1}
\end{figure}

The system studied in Ref.~\cite{dudovich2001transform} was $^{87}$Rb atomic vapor, whose energy levels and TPCA scheme are displayed in Fig.~\ref{f1}(a).  In this approach, it was considered a weak femtosecond pulse acting upon the system, exciting both transitions simultaneously. This is equivalent to make $E_u = E_l$ and $\tau = 0$ in Eq. (\ref{EQU Base}). Taking the laser frequency as $\omega_{u} = \omega_{l}$ and defining the detunings $|\omega_{ij} - \omega_l| = \Delta_{ij}$:

\begin{eqnarray}
    c_3(t) \approx - \frac{\mu_{12} \mu_{23}}{\hbar^2} \int^t_{- \infty} \int^{t'}_{- \infty} e^{i \Delta_{23} t'} e^{- i \Delta_{21} t''} \mathcal{E}_l^*(t')\mathcal{E}_l^*(t'') dt' dt" . \label{EQU Silb}
\end{eqnarray}

\noindent The final excitation after the pulse, $c_3(t \rightarrow \infty$), can be written in the frequency frame as

\begin{eqnarray}
    c_3 \approx \frac{\mu_{12} \mu_{23}}{i \hbar^2} \bigg[ i \pi \tilde{\mathcal{E}}^*(- \Delta_{23}) \tilde{\mathcal{E}}^*(\Delta_{12})
    \left. + \mathtt{p.v.} \int_{- \infty}^\infty d\omega \frac{ \tilde{\mathcal{E}}^*(\Delta_{12} - \Delta_{23} - \omega) \tilde{\mathcal{E}}^*(\omega) }{ \Delta_{12} - \omega } \right], \label{EQU Silb p.v.}
\end{eqnarray}

\noindent where p.v. is the principal value of Cauchy. Equation (\ref{EQU Silb p.v.}) highlights the phase-matching problem in resonant TPCA. The main issue is the second term, which integrates the off-resonance terms. Frequencies above the resonance have a phase shift of $\pi$ relative to the frequencies below the resonance. Then, for FTL pulses, this integral results in a destructive interference and limits the overall excitation \cite{dudovich2001transform}. Experimentally, the excitation to level 3 is observed through the fluorescence of its population $\rho_{33} = c_3^*c_3$. 

Analyzing Eq.(\ref{EQU Silb p.v.}), Dudovich et al. proposed in Ref.~\cite{dudovich2001transform} to use a Spatial Light Modulator (SLM) to fix the spectral phase mismatch of the off-resonance frequencies. In the corresponding experiment, a 778.1 nm laser, with linewidth $\Delta\lambda$ = 18 nm, interacted with an ensemble of rubidium atoms, exciting them to level 5$D$. To adjust the phase and improve the process, a 4 nm wide $\pi/2$ phase mask was applied on the laser beam between the resonant frequencies (780.2 nm and 776.0 nm). Under these conditions, Ref.~\cite{dudovich2001transform} reported $600\%$ increase in the outcome of the TPCA when compared to the same process driven by FTL pulses, a result well described by Eq.~(\ref{EQU Silb p.v.}). Prior to optimizing the TPCA using two distinct lasers, we will introduce and apply our adaptive optimization technique for this single pulse case, in order to test our optimization procedure against a well known problem.

\subsection{Genetic Algorithm}

For adaptive optimization, we employ a GA that modifies the pulse's spectral phase through a SLM in order to optimize the parameter of interest. In Secs.~\ref{SECSilb} and~\ref{SEC ThinSample}, the optimized quantity is the population $\rho_{33}$ of the excited state. The SLM which will apply a phase mask is simulated with 128 vectors, since this corresponds to a common commercial SLM model \cite{polycarpou2012adaptive, costanzo2016zero}. Each vector can apply a complex rotation of $[0, 2 \pi]$ in a specific frequency component. The spectral resolution of each vector is 1$\,$rad/ps, replicating the experimental conditions in \cite{polycarpou2012adaptive, costanzo2016zero}. Still, there are other factors to be considered about how a SLM applies spectral phase modifications in such experiment. In the Appendix we discuss these factors and how they were implemented. For the purpose of the GA, the rotation value of each vector in the SLM is written as 8 binary entries (8-bits), converted to decimal (float) only when a phase mask is applied. In this way it is possible to carry all the GA operations in binary format. For each run, the algorithm analyses 20 possible phase masks using the following steps:
\begin{itemize}
 \item[1.] After the application of a phase mask, the pulse is tested (in this work, it means ``integrated using a modeling equation"). 
 \item[2.] The result of the test is stored with the label of the phase mask. 
 \item[3.] After testing all 20 phase masks, the worst result is discarded and the best one is copied to fill the void. 
 \item[4.] Those 20 phase masks exchange information. This exchange is carried vector-wise, like a chromosome crossover, meaning that a pair of phase masks are chosen randomly. Each vector of the pair is cut at some binary entry point, and the information is swapped between vectors of that specific frequency. This happens for all vectors in the chosen pair of phase masks and each vector has its own cutting point. 
 \item[5.] After the exchange, random binary entries of all phase masks are chosen to have a random binary number printed on them. This concludes one GA iteration. 
 \item[6.] All this process is repeated over the next iteration. 
 \end{itemize}
\noindent
At the beginning of each run, all 20 GA phase masks are equal, having all vectors without any complex rotation in the pulse. The exchange rate among the 20 phase masks is 0.6 and the rate of random alterations on all the $20480$ ($20\times128\times8$) binary entries is 0.0003.

\begin{figure}
    \includegraphics[width = \linewidth]{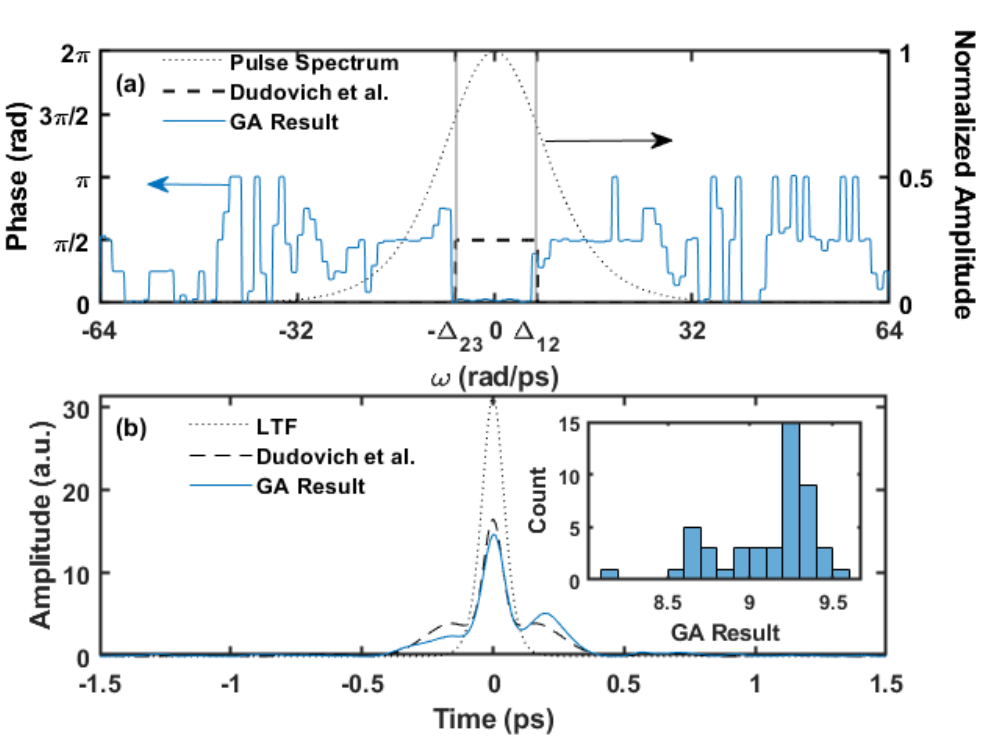}
    \caption{Genetic Algorithm optimized results for Eq. (\ref{EQU Silb}) and the parameters of Fig. \ref{f1}(a). Panel (a) shows the phase mask that achieved the best outcome, out of our 48 simulations, in solid blue. The dotted black curve is the pulse spectrum and the dashed black line provides the phase mask from Ref. \cite{dudovich2001transform}. The arrows point to the vertical axis of each curve. Panel (b) displays the resulting pulse shape, in solid blue, from the mask in (a). For comparison, in dashed black is the pulse resulting from the phase mask implemented in reference \cite{dudovich2001transform} and, in dotted black, the original FTL pulse. The inset shows the distribution of absorption optimization achieved by all 48 simulations.}
    \label{f2}
\end{figure}

\subsection{Phase Mask Optimization}

Now we present the GA results for optimization of $\rho_{33}$ from Eq. (\ref{EQU Silb}), with the parameters used in reference \cite{dudovich2001transform} and exhibited in Fig.~\ref{f1}(a). Figure~\ref{f2}(a) displays the best GA mask after 5000 iterations, in solid blue. The doted black curve is the pulse spectrum, to guide the reader in understanding how the phase shifts are related to each spectral component, and the dashed black curve is the phase window proposed by Ref. \cite{dudovich2001transform}. The GA found a solution similar to the one proposed in reference \cite{dudovich2001transform}, inserting a $\pi/2$ phase window between the resonant frequencies. The only difference was in how this phase window manifested itself. While in Ref. \cite{dudovich2001transform} it was proposed a $0 \rightarrow \pi/2 \rightarrow 0$ window, the GA came up with a $\pi/2\rightarrow 0 \rightarrow \pi/2$ window. Nonetheless, these phase windows are identical for this system, since only the absolute phase difference is taken into account. The random appearance of the mask in the spectral region far from the center is a result of the smaller contribution that these frequencies represent to the signal and the intrinsic difficult in the sequential excitation by single pulses. Since the resonance induces a phase mismatch of $\pi$ between frequencies above and below it, when we have two resonant frequencies from the same pulse it should be adjusted by $\pi$ after the first resonance and then by $\pi$ again after the second one. That is how Dudovich et al. improved the TPCA \cite{dudovich2001transform}. They used a $\pi/2$ phase step because the same field is accounted twice inside the integral in Eq. (\ref{EQU Silb p.v.}) and two $\pi/2$ phases add up, becoming $\pi$. On the other hand, the GA investigates the phase of each frequency randomly. So, for every change on one side of the spectrum, there is a compromise on the other, and the random pattern from the first changes prevails. To confirm the similarities between the two approaches, we compare pulse shapes in in Fig.~\ref{f2}(b). Notice how the GA output, in solid blue, is close to Ref.~\cite{dudovich2001transform}, in dashed black. They both can be compared to the FTL pulse, in dotted black. The inset brings the outcomes distribution of all 48 simulations we performed. Most notably is the predicted enhancement using our method. Our simulations achieved an average improvement of $9.3$ while the experiment from Ref.~\cite{dudovich2001transform} achieved $\sim 7$. Since we are working with an optimizing algorithm in a very sensitive system, it is expected some improvement when compared to an arbitrary solution. Also, this data demonstrates how GA results have a high tendency to be close to the best outcomes our method can provide. So, the chance of one in a set of three simulations, to have a result between $[9.2,9.6]$ is of $93\%$. This is a very useful information because it reliefs the need to perform too many simulations. With only three we have a high probability of finding a result that is very close to the best that GA can provide. With that, for now on we will work with three simulations, instead of 48, and focus on the best result among them.

\section{Two Lasers Problem}
\label{SEC ThinSample}

In this section we are going to investigate the situation depicted in Fig.~\ref{f1}(b), with two different coherent laser pulses exciting the sequential transition to level $5D$. We have first a signal laser pulse centered at 795 nm, with 7.5 nm bandwidth (FWHM), interacting resonantly with the transition $5S_{1/2} \rightarrow 5P_{1/2}$. Then the control pulse is centered at 762 nm with 10.4 nm bandwidth (FWHM), being resonant with the $5P_{1/2} \rightarrow 5D_{3/2}$ transition. This means that no pulse will have bandwidth to excite both transitions simultaneously, in contrast with last section. So, we do not need to worry about superposition of effects and matching the conditions from Ref.~\cite{carvalho2020enhanced}. $\tau$, in Fig.~\ref{f1}(b), represents the delay between the two pulses. The signal pulse is the only one going through a SLM. In this section this condition is merely cosmetic, but in the next section, when we approach the problem with a dense ensemble using the model proposed by Ref.~\cite{carvalho2020enhanced}, this will be necessary to respect the approximations taken for the model. For the sake of verification,  we also optimized the system considering a SLM modifying the control pulse as well as two SLMs, one on each pulse, but did not obtain any improvements or any additional difficulties in the GA response with respect to the data we will present here.

The system we are analyzing now is equivalent to make $\omega_{23} - \omega_u = \omega_{12} - \omega_l = 0$ in Eq. (\ref{EQU Base}), leading to

\begin{eqnarray}
    c_3(t) \approx - \frac{\mu_{12} \mu_{23}}{\hbar ^2} e^{i \omega_{23} \tau} \int_{-\infty}^{t - \tau} \int_{- \infty}^{t' + \tau} dt' dt" 
    \mathcal{E}_u^*(t') \mathcal{E}_l^*(t"). \label{EQU Thin}
\end{eqnarray}

\noindent Changing to the frequency frame for the field's envelopes and considering the long-time solution ($t \rightarrow \infty$), we can write

\begin{eqnarray}
    c_3 \approx \frac{\mu_{12} \mu_{23}}{i \hbar^2} e^{i \omega_{23} \tau} \Bigg[ i \pi \tilde{\mathcal{E}}_u^*(0) \tilde{\mathcal{E}}_l^*(0)
    \left. + \mathtt{p.v.} \int_{-\infty}^{\infty} d\omega \frac{\tilde{\mathcal{E}}_u^*(\omega) \tilde{\mathcal{E}}_l^*(- \omega) e^{- i \omega \tau}}{\omega} \right], \label{EQU Thin p.v.}
\end{eqnarray}

\noindent where p.v. is the principal value of Cauchy. Again, we have the $\pi$ phase difference between terms above and below the resonance inside the integral. However, now the pole is centralized with respect to the pulse spectrum and the integrand also involves an exponential term coming from the delay between the pulses, which gives a phase change directly related to $\tau$. The constant exponential outside the square brackets is irrelevant for $\rho_{33}$.

\subsection{Optimization Results}
\label{Opt2L}

\begin{figure}[ht]
    \includegraphics[width = \linewidth]{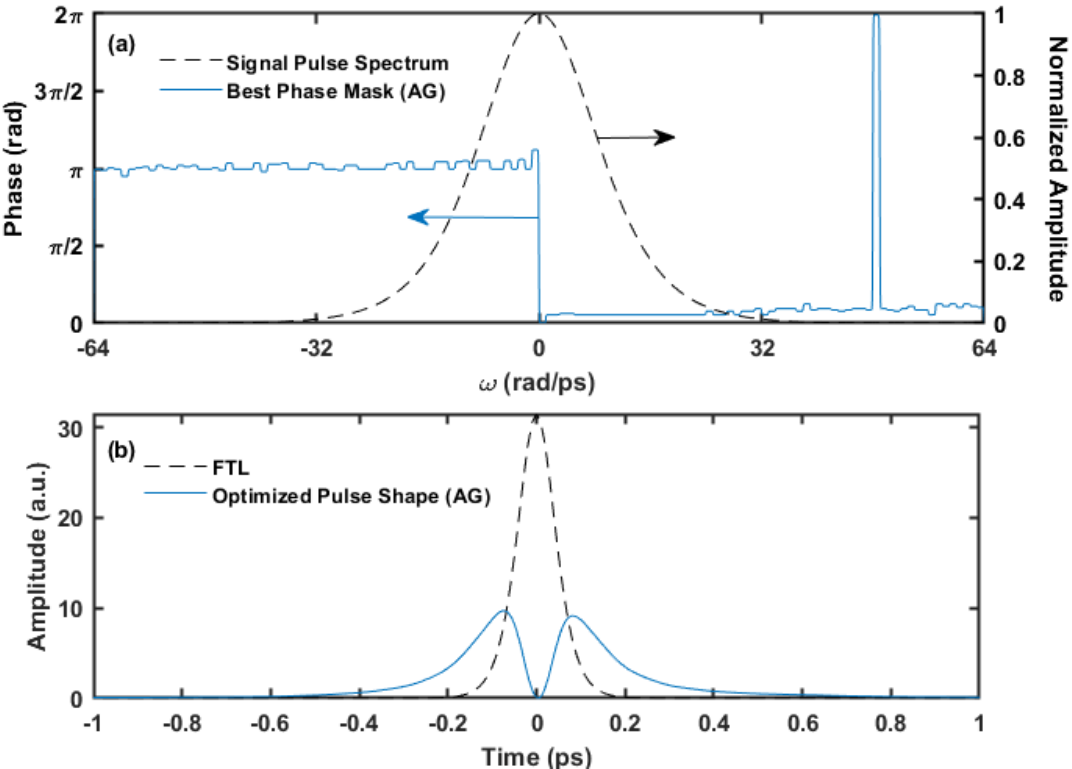}
    \caption{GA optimized results for Eq.~(\ref{EQU Thin p.v.}) with $\tau = 0$. Panel (a) presents the phase mask that achieved the best outcome, in solid blue. The pulse spectrum is shown by the dashed black line, for reference. The arrows point to the vertical axis of each curve. Panel (b) plots the pulse shape, in solid blue, resulting from the phase mask in (a). The dashed black curve is the initial FTL pulse. This spectral phase mask increased the absorption by 26 times.}
    \label{f3}
\end{figure}

The GA was employed to optimize of Eq.~(\ref{EQU Thin}), applying phase changes only to the $ \tilde{\mathcal{E}}_l(\omega)$ components. The most relevant result came from optimizing with $\tau$ fixed to zero, i.e. no delay between pulses. As seen in Fig.~\ref{f3}(a), the optimum phase mask has the appearance of a $\pi$ Heaviside function. In this case, the GA only acts over the oddity of the $1/\omega$ factor. Since the SLM modifies a single pulse inside the integral in Eq.~(\ref{EQU Thin p.v.}), differently from Sec.~\ref{SECSilb}, the phase adjustment is $\pi$ instead of $\pi/2$. This represents a better condition to fix the phase mismatch in the integrand of Eq.~(\ref{EQU Thin p.v.}). As a consequence, the phase mask has a better defined trace when compared to the optimum mask of Fig.~\ref{f2}(a).

The blue solid line in Fig.~\ref{f3}(b) plots the pulse shape resultant from applying the phase mask in panel (a). The dashed black curve is the original FTL pulse. At the end, all runs predicted an increase over 2600\%, an improvement far better than what we saw in Section (\ref{SECSilb}), indicating that the use of two lasers gives much better results than using a single one for the TPCA problem.

We performed $\tau$ scans to see if the delay between pulses was of any relevance, but the differences were minimal. The maximum fluorescence obtained was the same and the algorithm did not converge in any different manner. The phase mask changed only to compensate the delay, which created a periodic sawtooth structure superimposed to the Heaviside function. As a consequence, the signal pulse and the control pulse were once again reaching the atom simultaneously. In the end, $\tau$ proved itself irrelevant for the range of delay variations used for optimization in Ref.~\cite{carvalho2020enhanced}.

\subsection{Zero-area Pulses}
\label{SEC Zero Area}

So far we have been considering only the SLM between our initial FTL signal pulses and the atomic system. But in Sec. \ref{SECEnhanced} we will work with a dense atomic sample. The model of Ref.~\cite{carvalho2020enhanced} for dense ensembles concluded that only a fraction of the light path would contribute with TPCA, meaning that the signal pulse travels through a resonant medium before arriving at the interaction region, creating a $0\pi$-pulse (zero area pulse) before the TPCA. The area $\Theta$ of a pulse exciting a particular atomic transition is given by
\begin{equation}
    \Theta = \frac{1}{\hbar}\int^{\infty}_{- \infty} \boldsymbol{d} \cdot \boldsymbol{E} (t) dt,
\end{equation}
\noindent where $\boldsymbol{d}$ is the dipole moment of the transition and $\boldsymbol{E}$ is the envelope of the electric field. The area of a pulse provides directly the probability $P= \sin^2\left(\Theta /2\right)$ of finding a two-level atom in its excited state after the action of the pulse. The area theorem determines that any pulse with a small area in the entrance of the atomic sample consisting of an ensemble of two-level atoms will evolve to a pulse with zero area~\cite{Allen1987}. When the initial pulse has a bandwidth much larger than the atomic system, this zero area results from a distortion of the pulse's envelop, leading to some transient population in the excited state but zero excitation left after the action of the whole pulse~\cite{crisp1970propagation}. The common procedure to create zero-area pulses is to pass the light beam through an initially resonant atomic sample~\cite{costanzo2016zero,Dudovich2002}. 

\begin{figure}[ht]
    \includegraphics[width = \linewidth]{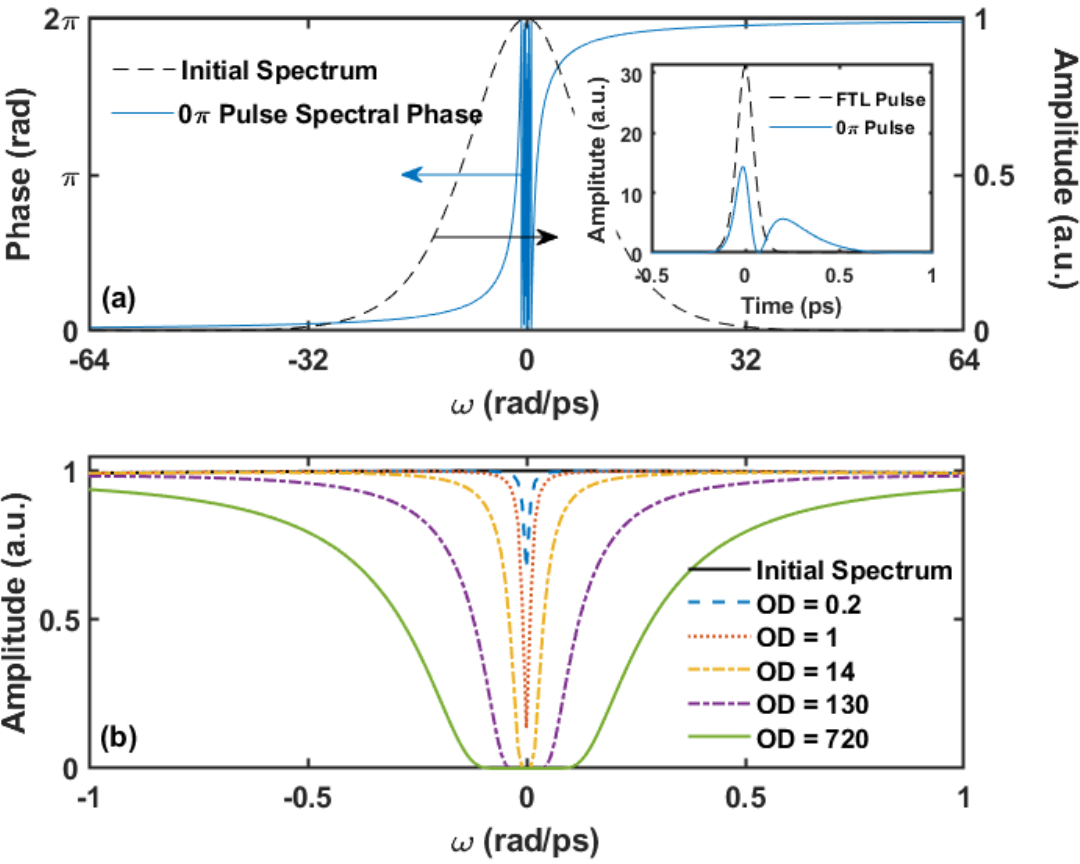}
    \caption{Typical characteristics of a zero-area pulse. (a) The blue solid line plots the spectral phase of the pulse leaving a sample of length $l$, described by the imaginary part of Eq.~(\ref{EQU Zero Area}). The atomic resonance ($\omega_a$) is set to zero. The modifications in the spectral phase can be compared to the dashed line, representing the pulse spectrum. The arrows point to the vertical axis of each curve. The inset is the shape of the pulse leaving the sample, in solid blue. The black dashed line is the shape of the FTL pulse before entering the atomic sample. (b) Spectrum of a $0\pi$ pulse, described by the real part of Eq.~(\ref{EQU Zero Area}), for different OD.}
    \label{f4}
\end{figure}

The expression describing the zero-area pulse leaving a sample of length $l$, in the frequency domain, is~\cite{crisp1970propagation,carvalho2020enhanced}
\begin{equation}
    \label{EQU Zero Area}
    E(\omega, l) = E(\omega, 0) e^{\frac{- \alpha_0 l}{1 - i (\omega - \omega_a) T_2}},
\end{equation}
\noindent with $\omega_a$ as the transition frequency, $\alpha_0$ the optical density, and $T_2$ the inhomogeneous time decay. The optical depth (OD) of the medium is given then by OD $= \alpha_0 l$. The zero-area pulse has various distinctive characteristics. First, its pulse phase follows a characteristic $S$ shaped curve when we limit the value between $[0,2\pi]$, with large oscillations around the atomic resonance, as shown in Fig.~\ref{f4}(a) with $\omega_a = 0$. As the optical depth increases, the oscillating region around resonance spreads to an increasing portion of the spectrum, as described by the imaginary exponential of Eq.~(\ref{EQU Zero Area}). The second characteristic is the oscillating pulse shape, seen in the inset of Fig.~\ref{f4}(a). The final profile is wider and has a lower peak amplitude, with the formation of smaller trail peaks. The sequence of peaks indicates the periods where the pulse's electric field changes sign, with subsequent periods acting in opposite directions with respect to the atomic excitation~\cite{carvalho2020enhanced}. The resultant spectrum is described by the real part of Eq.~(\ref{EQU Zero Area}) and can be seen in Fig.~\ref{f4}(b), where the spectral modifications provided by resonant media of different ODs are plotted.

\subsection{Optimization with Zero-area Pulses}
\label{SUBSEC Optimization with Zero-area Pulses}

\begin{figure}[ht]
    \includegraphics[width = \linewidth]{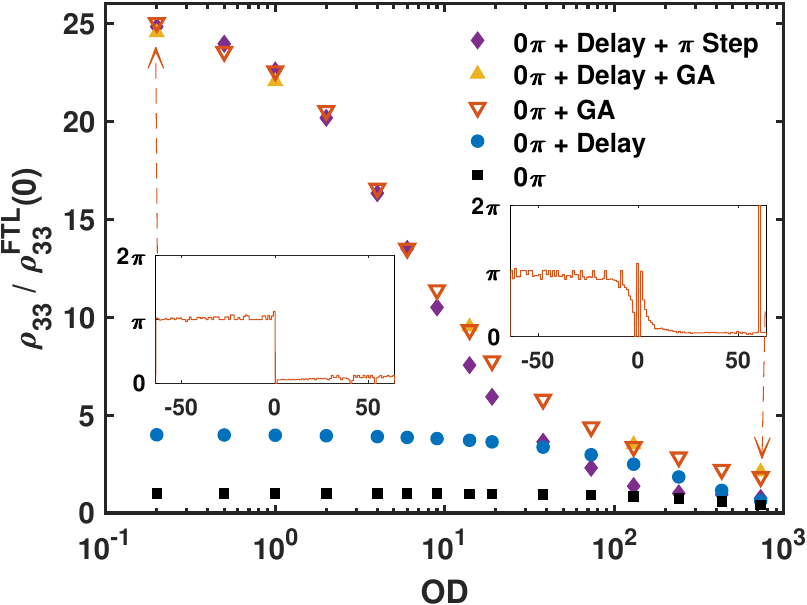}
    \caption{Impact of OD in the optimized two-photon cascade absorption with a zero-area pulse. The black squares are the results with a $0\pi$ pulse without optimization. The blue circles plot the absorption with delay ($\tau$) optimization alone. The red down triangles have only phase optimization ($\tau = 0$). The golden triangles are the results with phase and delay optimization. The purple diamonds are $0\pi$ pulses modified by a $\pi$ step phase with delay optimization. The normalization is relative to the absorption of a FTL pulse. The insets show two examples of the phase masks obtained by the GA for the downward triangles.}
    \label{f5}
\end{figure}

If we still consider a single atom, but now experiencing a pulse from the interaction region of a dense ensemble (meaning it will interact with a $0\pi$-pulse), it reveals some interesting behavior. In Fig.~\ref{f5} we display five results of a TPCA involving a $0\pi$ pulse acting in the lower transition, as a function of OD. The black squares plots the excitation induced in an atom by $0\pi$ pulses as we increase the OD of the medium that generated the $0\pi$ pulse. No optimization is applied for this curve. The vertical axis is normalized by the excitation induced by a FTL pulse. Notice how the induced excitation diminishes as we create a $0\pi$ pulse using higher OD. This behavior is explained by the loss in energy of the $0\pi$ pulses as they propagate through denser samples. So, one can look at Fig.~\ref{f5} in two ways: as an atom reacting to a modified pulse or as the maximum possible excitation induced by a pulse optimized for a particular section of an ensemble (when there is optimization).

If we apply some delay between the signal $0\pi$ pulse and the control pulse, in blue dots, we see that it is possible to obtain a substantial increase. For lower OD, it is possible to increase the excitation by four times, compared to a FTL pulse. For higher OD, the improvement is still solid, but decreases faster for ODs above 100, becoming almost negligible for OD = 720. We can understand this outcome by looking at Fig.~\ref{f4}(b) and Eq.~(\ref{EQU Thin p.v.}). In the frequency domain, the delay is a linear phase applied to the integral in Eq.~(\ref{EQU Thin p.v.}). This means that the interference will no longer be at its minimal (destructive) phase, but in some intermediate point. This alteration is responsible for increasing the excitation by four times for lower OD. What changes for higher OD is the absence of energy in the resonant region, where the interference is critical. Above OD = 10, as seen in Fig.~\ref{f4}(b), there is no energy left in the peak of resonance. From that point forward, we see a decrease in Fig.~\ref{f5}, becoming more evident above OD = 100, where the frequencies around the resonance are also becoming depleted of energy.

When we introduce GA optimization to the spectral phase, we see a remarkable improvement, in the red downward triangles. For low ODs, the tendency for $26\times$ improvement is clear, leading to what was commented in Sec. \ref{Opt2L} (which would be a case with OD = 0). When we compare the spectral phase optimization with the blue dots, that have only delay optimization, we see that it is much more sensitive to the variation of OD. Not only it has higher absolute variation, but it is also qualitatively different from the blue dots. While the blue dots decline faster above OD = 100, the downwards triangles have an intense decay around OD = 10, slowing down after OD = 100 to about the same pace (but slower) as the blue dots. This qualitative difference comes from the spectral optimization. While with the blue dots we had only a small spectral phase adjustment coming from the delay, when the GA seeks to improve the excitation it does so by solving the destructive interference problem completely. The GA solution gives access to all the energy around the resonance, which is responsible for increasing the excitation $26\times$ for OD = 0. But when we deal with a $0\pi$ pulse, it is depleted of energy in the resonant region. This situation is beyond any spectral phase adjustment, which explains why the GA fails to reach the same excitation levels when we increase the OD. Comparing to the sample delay optimization, while the decay of excitation seems similar between dots and downward triangles for OD above 100, the fully optimized pulses decay a little slower, since the spectral phase optimization will always deliver better results by construction. Ultimately, the GA results are able to stay above all outcomes of delay optimization alone, being $3\times$ better even in an extreme situation like OD = 720.

Once again, when we add both GA spectral phase and delay scan, there is no improvement in the excitation, as shown by the yellow triangles. We brought only a few samples to reinforce how the GA spectral phase optimization already accounts for any delay adjustments (withing the range of the SLM and experimental limitations). Just as before, any delay between pulses was compensated using a linear spectral phase whose $0 \rightarrow 2\pi$ rise rate in frequency matches the inverse of the delay in seconds. This confirms how delay and phase optimization are redundant in our analysis. We did continue to explore how the delay impacts optimization, but only came to the same conclusion. So, we will no longer consider delay optimization in this document.

Finally, if we were to take only a $\pi$ step spectral phase and combine it with a delay adjustment, would it be sufficient to solve the interference problem in a dense ensemble? This situation is displayed in purple diamonds in Fig.~\ref{f5}. We can see that a $\pi$ step solution offers the same outcome as the GA for OD below 10. But, passing that point, it is not able to achieve the same results due to increasing changes in the spectral phase originated from the $0\pi$ pulse modifications. For each of the purple diamonds we included a delay optimization to guarantee we are obtaining the best possible results (as one would do in an experiment). Still, we see that not only this is unable to keep the results at the same level as the GA after OD = 10, but it also becomes worse than the delay adjustment alone with $0\pi$ pulses, in blue dots, above OD$\approx 50$. This happens because the $0\pi$ spectral changes end up combining with the $\pi$-step phase to create a phase sum that is closer to the destruction interference when compared with the delay alone. The delay is unable to compensate the combination, resulting in less excitation for high OD in the purple diamonds. This is further evidenced by the insets in Fig.~\ref{f5}, where we can see two examples of optimizing phase masks for the downward triangles, one for OD = 19 which is very similar to what was obtained in Fig.~\ref{f3}(a) and another for OD = 720 that changes to compensate the phase demonstrated in Fig.~\ref{f4}(a). Notice that the GA results depend on the SLM resolution. For higher resolutions, the results would probably be better. However, we used the resolution from the experiment in Ref.~\cite{carvalho2020enhanced}.

\section{Enhanced absorption in a dense atomic medium}
\label{SECEnhanced}

Under the light of the previous sections, we move on to analyze the spectral phase optimization for absorption by a dense atomic medium as a whole. The system to be considered is composed by several of the individual atoms considered in Fig. \ref{f1}(b). This has important implications for the model and we will follow the considerations presented by Carvalho et al. for this system in \cite{carvalho2020enhanced}, since it demonstrated good correspondence with the experimental data. That said, we will briefly highlight the important aspects of the model here, but for more details, see \cite{carvalho2020enhanced}.

Considering our pair of pulses, propagating in the z-direction, delayed from each other by $\tau$, passing through an atomic sample with length $l$, the signal pulse coming out from the sample can be described in the frequency space as

\begin{equation}
    \label{Omega}
	\tilde{ \Omega }_s^{ (1) }(l, \omega) = \tilde{ \Omega }_s^{ (0) }(l, \omega) + \frac{ \Theta_c }{ 8 } e^{i \omega ( \tau + \frac{ l }{ c } ) } \int^{ \tau }_{ - \infty } dt F(\omega, t) ,
\end{equation}

\noindent where

\begin{eqnarray}
	F(\omega, t) = \frac{ 1 }{ \sqrt{ 2 \pi } } \int^{ \infty }_{ - \infty } d\omega' e^{ - i \omega' t} T( \omega, \omega' ) \tilde{ \Omega }^*_s (0, \omega') \tilde{ \Omega }^*_c (0, - \omega),
\end{eqnarray}

\noindent with

\begin{eqnarray}
		T(\omega, \omega') = \frac{ A(\omega) e^{ - OD[A(\omega) + A(\omega')] / 2} }{ [A(\omega) - A(\omega')] } \left\lbrace e^{ OD_c[A(\omega) - A(\omega')] / 2 }. - e^{ - OD_c[A(\omega) - A(\omega')] / 2 } \right\rbrace ,  \label{EQU Tww}
\end{eqnarray}

\noindent and

\begin{equation}
        \label{EQU Aw}
    	A(\omega) = \frac{ 1 }{ 1 - i \omega T_{2} }.
\end{equation}

\noindent $\tilde{ \Omega }_s^{ (1) }$ is the first order perturbation in $\Theta_c$ of the signal field Fourier transform, $\Theta_c$ is the control pulse area, $\omega$ is the angular frequency. $\tilde{ \Omega }_s^{ (0) }$ is the Fourier transform of the zero-area signal pulse, which represents no perturbation in $\Theta_c$. $\tilde{ \Omega }_c^*$ is the complex conjugate of the control field Fourier transform, $OD_{c}$ is the OD of the interaction region (at the center of the sample), and $T_{2}$ is the inhomogeneous time decay, given by $T_{2} = 1/(2\omega_{21}\sqrt{2ln(2)k_{b}T/m_{Rb}c^{2}})$, with $\omega_{21}$ as the frequency of emission from the excited to ground state, $k_{b}$ the Boltzmann constant, $T$ is the temperature in kelvins, $m_{Rb}$ is the rubidium mass and $c$ is the speed of light.

Equation (\ref{Omega}) shows that this pulse has two characteristics that we have discussed before. The first comes from $\tilde{ \Omega }_s^{ (0) }$, which is

\begin{equation}
	\tilde{ \Omega }_s^{ (0) }(z, \omega) = \tilde{ \Omega }_s(0, \omega)e^{ \left[ \frac{i \omega}{c} - \alpha_0 A(\omega) \right] z},
\end{equation}

\noindent the zero-area pulse in frequency space, discussed in Section \ref{SEC Zero Area}. The second characteristic comes from $e^{i \omega \tau}$, which will impose a linear variations of spectral phase in the pulse, like Eq. (\ref{EQU Thin p.v.}) in Sec.~\ref{SEC ThinSample}. Equation (\ref{Omega}) results from some approximations. An important approximation is the expansion in the control pulse area ($\Theta_c$), assuming that the control pulse keeps its area constant as it propagates through the interaction region inside the sample. This supposition about the control field led us to work with only one SLM, modifying the signal pulse, as mentioned in Sec.~\ref{SEC ThinSample}. Despite of some approximations, the results obtained in reference \cite{carvalho2020enhanced} shows a qualitative similarity between theory and experimental data. So, this theory will be used to anticipate the optimization of the experiment with the aid of the GA.

 The variation in the absorption of the pulse by the ensemble is written as 

\begin{equation}
	\Delta \alpha = \frac{U^{(0)}(l) - U^{(1)}(l)}{U^{(0)}(l)},
\end{equation}

\noindent where

\begin{equation}
U^{(i)}(l) = \int^{\infty}_{- \infty} d\omega \tilde{ \Omega }_s^{ (i) }(l, \omega).
\end{equation}

\noindent Ultimately, $\Delta \alpha$ is the quantity used inside the GA as the optimization parameter for this section.

\begin{figure}
    \includegraphics[width = \linewidth]{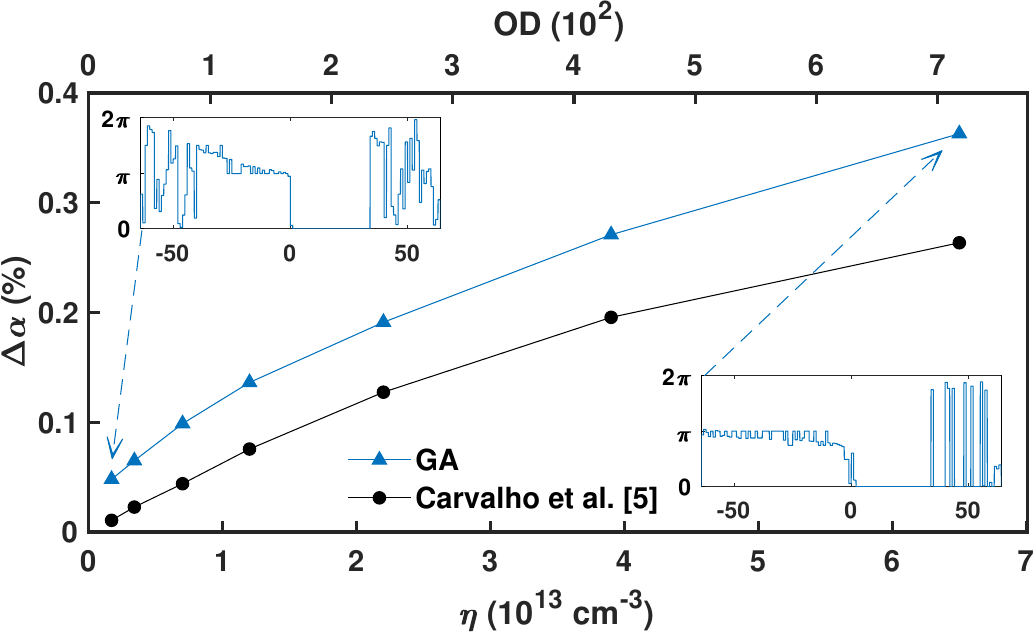}
    \caption{Optimization using Eq. (\ref{Omega}) for different values of OD. We compare the absorption for a FTL pulse (before the atomic sample), in black circles, and an optimized pulse, in blue triangles. The relative gain is higher for lower OD than for high OD. For OD = 19, $\Delta\alpha$ was $\sim 4$ times higher after the optimization. For OD = 720,  $\Delta\alpha$ was $\sim 1.5$ times higher after the optimization. The average relative improvement around all OD values is $\sim 2.5$. The OD values are translated to atomic density to make the comparison with the results from reference \cite{carvalho2020enhanced} easier. The insets show examples of the optimizing phase masks obtained by the GA.}
    \label{f6}
\end{figure}

\subsection{Optimized Absorption}
\label{SEC Optimized Ab}

The GA optimized the TPCA described by Eq.~(\ref{Omega}) using different OD values (or atomic density, $\eta$) on top of the best conditions demonstrated by Carvalho et al. \cite{carvalho2020enhanced}. In Fig.~\ref{f6}, we can see the GA performance optimizing the TPCA, in blue triangles, compared to the FTL-pulses results, obtained in reference \cite{carvalho2020enhanced}, in black circles. On average, the $\Delta\alpha$ was improved to slightly above twice the FTL value. But, as expected from the previous sections, the relative improvement is greater for lower OD values. This opens two possible scenarios for the experiment. One, where we can improve the absolute value of absorption by directly increasing the OD. Another, where we keep the OD low in order to improve the sensitivity to spectral phase, as also pointed out in Fig.~\ref{f5}. It is interesting to notice how the optimized absorption curve has a shape similar to the non-optimized one, with a tendency to reach a limit value as $\eta$ increases. The insets display the optimizing phase masks for two examples, both readily relatable to our previous results, demonstrating the efficiency of our analysis. The appearance of a sawtooth structure in these phase masks is caused by the compensation of the resonant medium before TPCA inside the sample, which acts like a delay line for the signal pulse.

\begin{figure}
    \center
    \includegraphics[width = 0.7\linewidth]{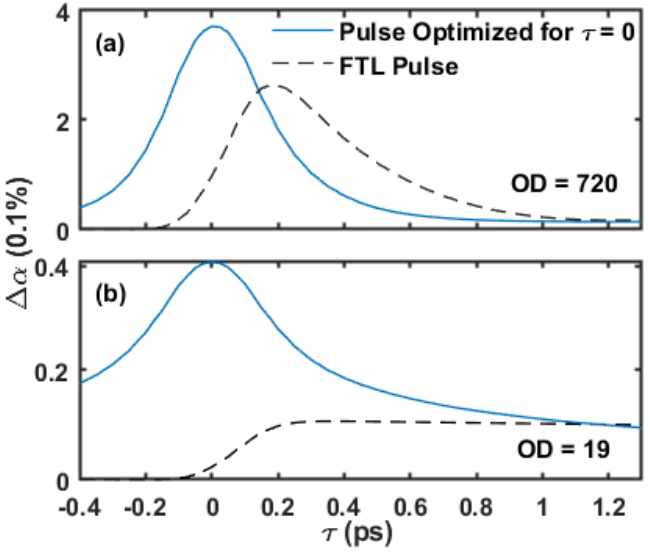}
    \caption{Absorption variation as function of pulse delay, $\Delta \alpha (\tau)$, for a given phase mask. The solid blue lines are the behavior for the optimized pulses. The dashed black lines are the results for FTL pulses, recalculated using the model provided by Ref.~\cite{carvalho2020enhanced}. We exemplify with the results of (a) OD = 720 and (b) OD = 19. These detailed curves demonstrate how the relative improvement is greater for lower OD.
    \label{f7}}
\end{figure}

Figure~\ref{f7} helps to understand the impact of the optimization for low and high OD. Here we see how the absorption is modified if, after choosing and applying an optimizing spectral phase mask, we scan the temporal delay ($\tau$) between pulses. The blue curves are the results for the optimizing phase masks, and with the peak of the curve being the value showed in Fig.~\ref{f6}. The dashed black curves are the results obtained by Carvalho et al. \cite{carvalho2020enhanced}. The plots illustrate how the relative improvement after the spectral phase optimization is higher for lower OD. One final aspect to be mentioned is the location of the peak in the blue solid curves at $\tau = 0$. As we discussed before, the GA dismissed the need of any delay (and it was tested), contrasting with the dashed black curves.

\section{Conclusion and Prospects}
\label{conclusion}

We demonstrated how to optimize simple TPCA to $2600\%$ enhancement with spectral phase optimization, using a GA as an aid to understand the sources of improvement to the point of dismissing the use of delay adjustments. We also had positive results with the TPCA modeled in reference \cite{carvalho2020enhanced}, showing it can indeed be optimized by spectral phase. Despite the complexity of the parameters and model, the observation of the spectral phase optimization for the model of an individual atom can anticipate some of the optimization for a complex system in a simplistic way. With this approach we demonstrated an average relative enhancement for TPCA in a dense medium twice above what was previously demonstrated. Besides that, our results are ready to be tested experimentally.

Our results also point that we have reached some limits of the spectral phase improvement for the considered SLM. We ran the GA many times, always until it was saturated, and all the runs led to the same results. This makes us believe there are no alternative routes to improve these processes through spectral phase adjustment under the experimental constraints we imposed. Nevertheless, the GA itself can be improved. For example, it can be faster for some experiments, using modifications weighted by the spectrum instead of random ones; or simulated with a SLM of higher resolution; or maybe combined with a frequency cut like in reference \cite{dudovich2001transform}. These changes, even if unable to greatly improve what we presented here, might make the technique more suitable for different applications.

\section*{Appendix}
\label{Appendix}

For the simulations, we defined the following parameters: phase mask spectral efficiency, time window size and temporal resolution. The phase mask spectral efficiency is limited by the spectral resolution. Since we want to compare these simulations with an experiment, we used the resolution of 0.5 rad/ps, from a diffraction grating available. We also had to take into account that the SLM is discrete and its efficiency varies from the edges to the center. So, we smooth out its limitation in resolution for each frequency entry in the phase mask, and we used the spectral resolution as a factor in the phase mask spectral efficiency. So, a logistic function was chosen, where it is given by:

\begin{equation}
    f(x) = \frac{M}{1 + e^{-A(x - x_0)}},
\end{equation}

\noindent where $M$ is the maximum function value, $A$ is the curve inclination and $x_0$ is where the function reaches the value $M/2$.

\begin{figure}
    \includegraphics[width = \linewidth]{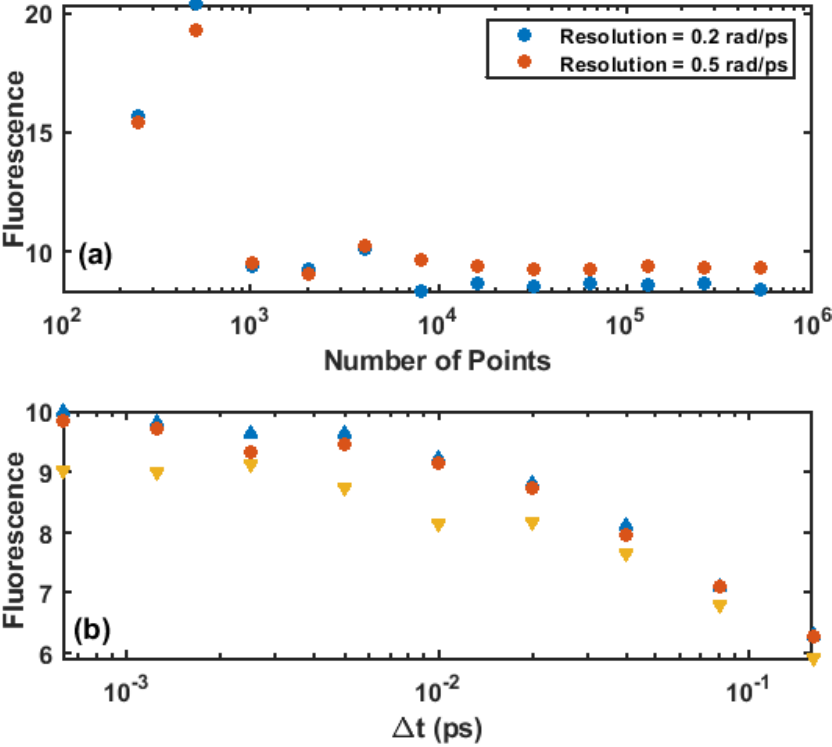}
    \caption{(a) AG result versus the number of points in the simulation. The blue circles are simulations with higher (0.2 rad/ps) spectral resolution than the red circles (0.5 rad/ps). (b) AG result versus temporal resolution. Time window of $\sim$320 ps. For each configuration we made three simulations. The blue triangle is the best result, the yellow inverted triangle is the worse and the red circle is the complementary one.
    \label{f8}}
\end{figure}

The time window considered has to be sufficient to run the simulations correctly. For the simulations in Fig.~\ref{f2}, we tested the numerical stability by varying the number of points. The results are shown in Fig.~\ref{f8}(a). A minimum number of points is required by the solution, to stay in the same region. Small fluctuations are expected due to the AG probabilistic nature. In Fig.~\ref{f2} $\sim$ 32000 points were used. For any variation in the parameters of simulation, it is necessary to verify the stability again.

The last parameter is the temporal resolution. In this case, the time window was fixed at $\sim$ 320 ps and the temporal resolution was changed. To keep the time window at 320 ps, it is necessary to increase the number of points proportionally to the temporal resolution. The result is shown in Fig.~\ref{f8}(b), which indicates the importance of keeping the temporal resolution below 0.01 ps. Thus, we used the temporal resolution of 0.005 ps in Fig.~\ref{f2}. The temporal resolution must be tested every time the problem conditions change.

\section*{Funding}

This work was supported by the Brazilian funding agencies CNPq (program INCT-IQ, No. 465469/2014-0), CAPES (program PROEX 534/2018, No. 23038.003382/2018-39), FAPESP (2021/06535-0), and FACEPE (program PRONEM 08/2014, No. APQ-1178- 1.05/14).

\section*{Disclosures}

The authors declare no conflicts of interest.

\bibliographystyle{plain}
\bibliography{mylib.bib}

\end{document}